\documentclass{Rinton-P9x6}
\usepackage{latexsym} 

\begin{document}

\title{Low-energy effective action in extended supersymmetric gauge theories.}

\author{A.T. Banin}

\address{Institute of Mathematics, Novosibirsk, 630090, Russia\\
}

\author{I.L. Buchbinder}

\address{Department of Theoretical Physics, Tomsk, 634041, Russia \\
E-mail: joseph@tspu.edu.ru}

\author{N.G. Pletnev}

\address{Institute of Mathematics, Novosibirsk, 630090, Russia\\
E-mail: pletnev@math.nsc.ru}


\maketitle

\abstracts{ We briefly review a recent progress in constructing
the low-energy effective action in ${\cal N}=2,4$ super Yang-Mills
theories. Using superfield methods we study the one- and two-loop
contributions to the effective action in the Coulomb and
non-Abelian phases.  General structure of low-energy corrections
to the effective action is discussed.}

\section{Introduction}
Supersymmetric field theories possess many remarkable properties
both in the classical and in the quantum levels. The supersymmetry
imposes rigid restrictions on a structure of quantum corrections.
In some cases these restrictions can be so strong that they allow
one to obtain exact results for the effective action at low
energies. In ${\cal N}=1$ SUSY models the supersymmetry
requirements lead to the known non-renormalizations theorems (see
e.g. \cite{ide}) and can provide an exact non-perturbative
determination of the chiral potential \cite{seib}.

It is evident that the more extended supersymmetry presents in the
model the more strong restrictions are imposed on the effective
action. In ${\cal N}=2$ SYM theories supersymmetry requirements
(together with duality) allow to get the exact solution for the
holomorphic part of the effective action \cite{sewi}. In ${\cal
N}=4$ SYM theory the supersymmetry and the superconformal
invariance provide finiteness of the theory and fix an exact form
of the non-holomorphic potential which gives the leading
low-energy contributions to effective action in ${\cal N}=2$
vector multiplet sector \cite{dise}. Generalization of this
non-holomorphic potential is the exact complete low-energy
effective action depending on all fields of the ${\cal N}=4$
vector multiplet \cite{bi}.

We consider the effective action on the base of superfield
formulations of extended supersymmetric models in ${\cal N}=1$
superspace and ${\cal N}=2$ harmonic superspace. Use of harmonic
superspace formulation gives a possibility to explore manifest
${\cal N}=2$ supersymmetry. However, since the operator techniques
leading to supersymmetric generalizations of the Heizenberg-Euler
or the Schwinger effective Lagrangians are still well developed
only in ${\cal N}=1$ superspace, we use ${\cal N}=1$ superspace
approach for construction of the effective actions beyond leading
low-energy approximation and on non-Abelian background.

\section{${\cal N}=4$ SYM effective action: exact low-energy effective action depending
on all fields of ${\cal N}=4$ vector multiplet and two-loop effective action
in sector of ${\cal N}=2$ vector multiplet}

In this section we briefly review a recent progress in
construction of the ${\cal N}=4$ SYM low-energy effective action
for the Coulomb phase in the framework of the ${\cal N}=2$
harmonic superspace formulation \cite{gikos}.

The harmonic superspace approach \cite{gikos} was successfully
used to study the effective action \cite{bpts}, \cite{bk},
\cite{bbiko}, \cite{bk1}, \cite{mckuz}. The main attractive
feature of such an approach is the possibility to preserve a
manifest ${\cal N}=2$ supersymmetry on all steps of quantum
calculations. For ${\cal N}=2$ SYM models in the harmonic
superspace the background field method was developed in the papers
\cite{bbko0}. Exploring the hidden ${\cal N}=2$ supersymmetry of
the ${\cal N}=4$ SYM theory formulated in the ${\cal N}=2$
harmonic superspace, the non-holomorphic potential can be
explicitly completed by the appropriate hypermultiplet-dependent
terms to the entire ${\cal N}=4$ supersymmetric form. Direct
calculation in ${\cal N}=2$ harmonical superspace allowed to
obtain as the exact form of the non-holomorphic potential
\cite{bk1} as the corresponding hypermultiplet dependent
complement \cite{bi}, \cite{bip}
\begin{equation}\label{1}
\Gamma[{\cal W},\bar{\cal W},q^+] = c\int d^{12}z\,
\left[\ln({\cal W})\ln(\bar{\cal W})+ {\cal L}_q({\cal
W},\bar{\cal W},q^+)\right]~,
\end{equation}
with function
\begin{equation}\label{2}
{\cal L}_q ({\cal W}, \bar{\cal W}, q^+)= \left(
(X-1)\frac{\mbox{ln}(X-1)}{X} + [\mbox{Li}_2(X)-1] \right)~,
\label{21}
\end{equation}
here $X=\left(-{q^{ia}q_{ia}\over {\cal W}\bar{\cal W}}\right)$;
$q^{ia}$ is the hypermultiplet superfield (see details of
denotations in \cite{bi}); ${\rm Li}_{2}(X)$ is the Euler
dilogariphm function. The bosonic component of the effective
action corresponding to (\ref{1}, \ref{2}) looks like
$F^4/(|\phi|^2 + f_{ia}\,f^{ia})^2$ where $\phi$ is the complex
scalar from ${\cal N}=2$ vector multiplet and $f_{ia}$ are the
scalars from hypermultiplet (see the details in \cite{bi}). The
effective Lagrangian (\ref{21}) was firstly found on the base of
purely algebraic analysis \cite{bi} and then reproduced by quantum
field theory calculations using ${\cal N}=2$ background field
method and the harmonic supergraphs technique.

Study of the two-loop structure of the ${\cal N}=4$ SYM effective
action for $SU(N+1)$ gauge group spontaneously broken down to
$SU(N)\times U(1)$ has been undertaken in the work \cite{bpts} to
clarify a possibility to describe D3-branes interactions in the
superstring theory in the terms of the effective action in the
${\cal N}=4$ SYM theory. In particular, in the large $N$ limit in
case of $U(1)$ constant background the ${\cal N}=2$ superconformal
invariant two-loop contribution to the effective action,
containing $F^6$-term in its component form, has been calculated.
It was shown that the two-loop effective action in the ${\cal
N}=2$ vector multiplet sector includes the following term
\begin{equation} \label{3}\Gamma_{(2)}
= N^2 g^2\frac{1}{3\cdot16(4\pi)^4}\int d^{12} z\
(\frac{1}{\bar{\cal W}^2} \ln\frac{\cal W}{\mu} {\cal
D}^4\ln\frac{\cal W}{\mu}+h.c.)
\end{equation}
Namely this functional leads to $F^6$ term in components. It was
proved that both the coefficient at one-loop $F^4$ term and the
coefficient at two-loop $F^6$ term in ${\cal N}=4$ SYM effective
action exactly correspond to the corresponding coefficients of the
Born-Infeld action expansion in the supergravity background (see
the details in Ref. \cite{bk} for the one-loop effective action
and in Ref. \cite{bpts} for two-loop effective action). It should
be pointed out the new covariant approach to study of one- and
two-loop contributions to superfield effective action for ${\cal
N}=2,4$ SYM theories \cite{mckuz}.

\section{The one-loop effective action in ${\cal N}=2,4$ SYM theories beyond leading
low-energy approximation}

In this section we briefly review a recent progress in studying
the one-loop ${\cal N}=2$ SYM theory for Abelian and non-Abelian
backgrounds and for ${\cal N}=4$ SYM effective action beyond of
leading low-energy approximation \cite{pbb1}, \cite{pbb2},
\cite{pbb3}.

We consider a hypermultiplet model coupled to external Abelian
${\cal N}=2$ vector multiplet using ${\cal N}=1$ superfield
formulation and study the induced effective action for ${\cal
N}=2$ vector multiplet. Non-holomorphic contributions to the
effective action are written as a sum of three terms. First of
these terms is
\begin{equation}
\left(\Gamma_{W\bar{W}}\right)_{\rm fin}= {1 \over (4\pi)^2}\int
d^8z\,\int_{0}^{\infty} dt\,t{\rm e}^{-t} {W^2 \bar{W}^2 \over
(\Phi\bar{\Phi})^2}\zeta(t\bar{\Psi}, t\Psi)~,\label{wwfin}
\end{equation}
where the function $\zeta(x,y)$ was defined in \cite{pbb1} and
quantities $\Psi$, $\bar{\Psi}$ are scalars with respect to ${\cal
N}=1$ superconformal group.

The other two terms are obtained one from another by the replacement
$ \left(\Gamma_{\Phi\bar{\Phi}}^{+}\right)_{\rm fin}=
\left(\Gamma_{\Phi\bar{\Phi}}^{-}\right)_{\rm fin}
(\Psi\leftrightarrow \bar{\Psi})$ and
\begin{eqnarray}
\left(\Gamma_{\Phi\bar{\Phi}}^{-}\right)_{\rm fin}&= & {1 \over 4
(4\pi)^2}\int d^8z\, \int_{0}^{\infty} {dt \over t^2}{\rm e}^{-t}
\Phi\bar{\Phi}\,\xi(t\bar{\Psi},t\Psi) - \nonumber\\
&-& {1 \over 12 (4\pi)^2}\int d^8z\,\int_{0}^{\infty} dt\,t{\rm
e}^{-t} {W^2\bar{W}^2 \over (\Phi\bar{\Phi})^2}\,
\lambda(t\bar{\Psi}, t\Psi)\tau(t\bar{\Psi},t\Psi),\label{gp}
\end{eqnarray}
where $\lambda(x,y)~,\, \xi(x,y)~, \tau(x,y)$ are some functions
found in \cite{pbb1}. One can show that the functionals
(\ref{wwfin}, \ref{gp}) can be rewritten in manifestly ${\cal
N}=2$ superconformal invariant form.

Now we consider a structure of the effective action of ${\cal
N}=2$ SYM model in a non-Abelian phase. We formulate the model in
${\cal N}=1$ superspace, use the background field method and
impose the gauge-fixing conditions depending on the gauge
parameters $\alpha$, $\lambda$ and $\bar\lambda$ (see the details
in \cite{pbb2}, \cite{ow}).

The gauge-dependent contribution is concentrated in the
non-holomorphic potential ${\cal H}$ and can be found at any fixed
choice of gauge parameters. For the Landau-DeWitt gauge, i.e. then
$\alpha =0$, $\lambda=\bar\lambda=1$ we obtain \cite{pbb2}
\begin{eqnarray}
&2(4\pi)^{2}{\cal H}= \ln (2)\ln (1-s^{2})\,+ {1\over
\sqrt{2}}\ln \left({\sqrt{2}-1\over \sqrt{2}+1}\right)
\ln (1-s^{2})-{\rm Li}_{2}\left({s^{2}\over 2}\right)+ & \nonumber\\
&+{\sqrt{2}-1\over\sqrt{2}} \left[{\rm
Li}_{2}\left({s-1\over\sqrt{2}-1}\right)+ {\rm
Li}_{2}\left(-{s+1\over\sqrt{2}-1}\right)\right]+
{\sqrt{2}+1\over\sqrt{2}} \left[{\rm
Li}_{2}\left({s+1\over\sqrt{2}+1}\right)+ {\rm
Li}_{2}\left({1-s\over\sqrt{2}+1}\right)\right],& \label{gdep}
\end{eqnarray}
where the notations $ s^2 = 1-{\Phi^{2}\bar{\Phi}^{2}\over
(\Phi\bar{\Phi})^{2}} < 0$,  $t={{\Phi}\bar{\Phi}\over
\sqrt{{\Phi}^{2}\bar{\Phi}^{2}}}$ are used; ${\rm Li}_{2}(X)$ is
the Euler dilogariphm function. As we see, the form of
non-holomorphic potential, in general, depends on a gauge choice.
This fact can lead to the ambiguous in derivative expansion in
non-Abelian phase. Analogous problem also arises when one defines
non-Abelian Born-Infeld action \cite{tsey}.

Now we consider a problem of the hypermultiplet completion to the
next-to-leading terms $F^8, F^{10},...$ for ${\cal N}=4$ SYM
theory \cite{pbb3}. Our aim is to develop a systematic procedure
allowing to construct an expansion of the one-loop effective
action in a power series of Abelian strength $F$. It was shown
\cite{bk}, \cite{pbb3} that the one-loop contribution can be
written as a power expansion of dimensionless combinations $ {\bar
\Psi}^2 = \frac{1}{M^2} \,\nabla^2 {W^2}$, $\Psi^2 =\frac{1}{M^2}
\,{\bar \nabla}^2 {\bar W}^2$. The quantity $M$ depends on the
chiral fields, which contain scalar fields from the ${\cal N}=2$
vector multiplet and the hypermultiplet. In the constant field
approximation this expansion is summed to the following expression
for the whole one-loop effective action (see details in
\cite{bk}):
\begin{equation}\label{n4gamma}
\Gamma = \frac{1}{8\pi^{2}}\int{\rm d}^8 z \int_{0}^{\infty}{\rm
d}t\,t\,{\rm e}^{-t}\frac{ W^{2}\bar{W}^{2}}{M^2}\,\omega (t\Psi,
t\bar{\Psi})~,
\end{equation}
where function $\omega$ was defined in \cite{bpts}. The difference
between the effective actions with and without the hypermultiplet
background fields hides in the quantity $M$ \cite{pbb3}. The
expansion of the function $\omega$ in power of $\Psi, \bar{\Psi}$
leads to the the series for the effective action (\ref{n4gamma}):
\begin{equation}\label{gdecom}
\Gamma = \Gamma_{(0)}+\Gamma_{(2)}+\Gamma_{(3)}+\cdots,\quad
\Gamma_{(n)} \sim \sum_{m + l =
n}c_{m,l}\Psi^{2m}\bar{\Psi}^{2l}~.
\end{equation}
In the bosonic sector, this expansion corresponds to expansion in
powers of the strength $F$, namely $\Gamma_{(n)}\sim
F^{4+2n}/M^{2+2n}$. The calculations of $\Gamma_{(0)}$ lead to the
expression, which was firstly found in \cite{bi}, \cite{bip}. The
${\cal N}=2$ form of next term ($\sim F^8$) in the series
(\ref{gdecom}) is reconstructed to the following expression for
$\Gamma_{(2)}$:
\begin{eqnarray}\label{g2}
&\Gamma_{(2)}= {1\over 2\cdot5!(4\pi)^{2}}\int d^{12}z
{\bf\Psi^{2}\bar{\Psi}^{2}}({1\over (1-X)^{2}}+{4\over (1-X)}+&
\nonumber\\ &+{6X-4\over X^{3}}\ln (1-X)+4{X-1\over X^{2}})~,
\end{eqnarray}
here ${\bf\Psi}^{2}= {1\over {\cal W}^2} \bar{D}^4\ln \bar{\cal
W}$. This relation defines ${\cal N}=2$ superfield form of $F^8$
contribution to the effective action depending on all fields of
${\cal N}=4$ vector multiplet. Moreover, in the paper of Refs.
\cite{pbb3} it was shown that any term in (\ref{gdecom}) can be
written in terms of on--shell ${\cal N}=2$ superfields.

\section{Conclusion}

We have presented the recent results on a structure of the
low-energy  effective action in extended supersymmetric field theories
obtained in our papers \cite{bi}, \cite{bpts}, \cite{bk},
 \cite{bk1}, \cite{bip}, \cite{pbb1}, \cite{pbb2}, \cite{pbb3}. The low-energy effective action has
been studied using the the superfield formulations of these
theories in standard ${\cal N}=1$ superspace and the ${\cal N}=2$
harmonic superspace.

Exact low-energy effective action depending on all fields of the
${\cal N}=4$ vector multiplet has been constructed for ${\cal
N}=4$ SYM theory in the Coulomb phase. This result has been
firstly obtained by analyzing the invariance of the effective
action under hidden ${\cal N}=2$ supersymmetry transformations in
${\cal N}=2$ harmonic superspace \cite{bi} and then reproduced by
direct harmonic supergraph calculations \cite{bip}. The two-loop
effective action in ${\cal N}=2$ vector multiplet sector was
studied \cite{bpts} and it was proved that in the t'Hooft limit
the coefficient at $F^6$ term exactly coincides with one in the
Born-Infeld action.

The one-loop effective action of various ${\cal N}=2$
supersymmetric models including ${\cal N}=4$ SYM theory has been
studied in the Coulomb and non-Abelian phases taking into account
dependence both on the fields of ${\cal N}=2$ vector multiplet and
hypermultiplet \cite{pbb1}, \cite{pbb2}. New ${\cal N}=1$
covariant and gauge invariant procedure for finding the effective
action was formulated and a derivative expansion was developed on
its basis. The concrete results are: the effective action of the
${\cal N}=2$ vector multiplet induced by the hypermultiplet, gauge
dependence of the effective action on a non-Abelian background in
${\cal N}=2$ SYM theory and the one-loop effective action
including dependence on all powers of the Abelian strength and all
powers of hypermultiplet fields in ${\cal N}=4$ SYM theory. In the
leading order this action reproduces the complete ${\cal N}=4$
supersymmetric low-energy effective action found in \cite{bi} and
allows to get a higher order correction containing the terms $F^8,
F^{10}, \ldots$ with the corresponding hypermultiplet completions.

\section*{Acknowledgments}
I.L.B is very thankful to the Organizing Committee of QFEXT03 and
especially Prof. K. Milton for kind support of his participation
at the conference. The work was supported in part by INTAS grant,
INTAS-00-00254 and RFBR grant, project No 03-02-16193.  I.L.B is
grateful to grant for Leading Russian Scientific Schools, project
No 1252.2003.2 for partial support. The work of N.G.P and A.T.B
was supported in part by RFBR grant, project No 02-02-17884.

\end{document}